\documentclass[12pt]{article}
\usepackage{amsmath}
\usepackage{epsf}
\usepackage{graphicx}
\usepackage{amsfonts}

\begin{document}
\newcommand {\artitle}
{\textbf{\large{Design and Development of Low Cost PC Based Real Time Temperature and Humidity Monitoring System}}}
\newcommand {\arauthor}
{ Nungleppam Monoranjan Singh\footnote{\it{E-mail:} mranjansn@yahoo.com}
 and Kanak Chandra Sarma\footnote{
  \it{E-mail:} kanak\_sarma50@rediffmail.com}}

\newcommand{\araddress}
{Department of Instrumentation \& USIC,\\Gauhati University, Guwahati-781014,India}
\newcommand{\arabstract}
 {This paper presents the design and development of a low cost Data Acquisition System (DAS) using PIC12F675 microcontroller for real time temperature and humidity monitoring. The designed DAS has 4 analog input channels having 10-bit resolution and was interfaced through the serial port of the PC. A precision integrated temperature sensor and an instrumentation-quality RH (Relative Humidity) sensor were used for sensing the temperature and humidity respectively. The firmware was written in Basic and compiled using Oshonsoft PIC IDE and downloaded to the microcontroller by using PICkit2 programmer. An application program was also developed using Visual Basic 6, which allows displaying the waveform of the signal(s) in real time and the data can be saved into the hard disk of the computer for future use and analysis. It can also be interfaced to the USB port of the PC or laptop using USB to serial adapter BAFO BF-810. Thus, the designed low cost device works with the legacy hardware as well as the modern USB interface.\\

\textbf{ Keywords:} Data Acquisition System, Microcontroller, Serial Port, USB interface, 10-bit resolution, Temperature, Relative Humidity\\}

\begin{titlepage}
\begin{center}
{\large{\bf\artitle}}
\bigskip
\bigskip

{\arauthor}
\\
\bigskip
\mbox{}
{\it \araddress}
\\

\vspace{.5in}
{\bf Abstract} \bigskip \end{center}\setcounter{page}{0}
\arabstract
\end{titlepage}

\section{Introduction}
It is very much essential in laboratories as well as in industries to monitor temperature and humidity continuously. Earlier measurements had been done through manual measurements from analog instruments, such as thermometers, manometers and hygrometers. Such type of measurements can’t fulfill the current requirements in terms of the time duration and accuracy. The efficient solution for this problem is to develop a data logger [\ref{1},\ref{2}]. Microcontroller based embedded design for monitoring temperature has brought a revolutionary change [\ref{3},\ref{4},\ref{5}]. The standalone system allows monitoring and controlling the physical parameters but, not suitable to log data for a long duration of time. Now a days, PC becomes very common, cheap and reliable. Further development in data logging took place as people begin to create PC-based data logging systems [\ref{6}]. A low cost PC based real time data logging system can be used for such purpose. This will allow the custom control, display and storing of the recorded data to the PC.\\

In most of the universities in the developing countries, the number of DAQ (Data Acquisition) cards in their laboratories are inadequate due to the high prices of the ready made DAQ cards. So, researchers are trying to develop customized, low cost DAS to meet their needs [\ref{7}]. Since 1995, where computer technology were not so advanced as compared with to days generation, researchers have been designing PC-Based DAS to be low cost, high speed response, minimized sample skew, compatibility etc. [\ref{8}]. New trends in the design of DAS is the flexibility, customized firmware and software platform for rapid development of PC based DAS [\ref{9}]. Design, monitoring and controlling of PC based system plays an important role in the present technology of 21st century. Such type of developments are  applied in various industries for PC based process control, green house monitoring and control system, agriculture and food industries, etc [\ref{10},\ref{11}].\\

The present work is to explore the possibility of continuously monitoring and storing the temperature and humidity through PC. The designed system is connected to the PC through RS-232 serial port. Even though there are many readymade instruments available, still research is going on to make a better, reliable and cost effective instrument. Today, many signal acquisition systems are PC based because of their high efficiency. There are many data acquisition systems available in the market for acquiring data and analysis. They have a large number of functions which are not required for the application and is also expensive. Thus, a custom cost effective hardware is designed, and the corresponding application software and firmware are also developed.

\section{Materials and Methods}

The circuit diagram of the experimental setup is shown in Fig. 1. The hardware and software description of the system for the real time monitoring of temperature and humidity are described below.

\subsection{Data acquisition unit}

A data acquisition system (DAS) has been developed using PIC12F675 microcontroller. PIC12F675 is a mid range microcontroller having an internal crystal oscillator. Fig. 1 shows the circuit diagram of the designed system. It consists of a +5 volt regulated power supply, PIC12F675 microcontroller
\begin{figure}[!h]
\centering \includegraphics[width =12 cm]{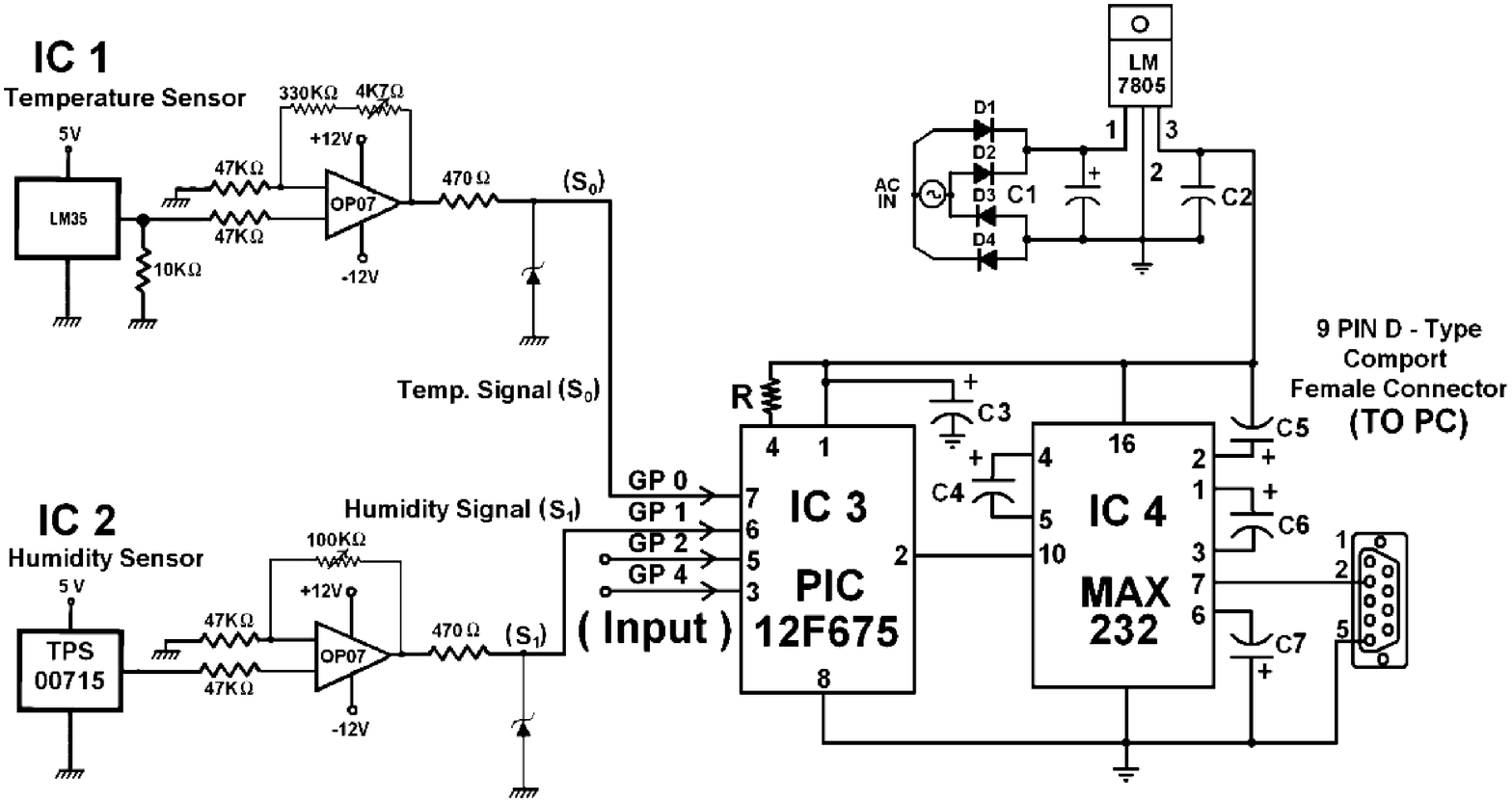}
\caption{Circuit diagram of the experimental setup}
\end{figure}
 and a MAX232 (RS-232 driver). PIC12F675 has four analog input channels having 10 bit resolution ADC, in which the entire operation is controlled by the firmware. Pin 2 of the PIC12F675 is used to send the serial data to the PC (Personal Computer) through the MAX232 driver IC. The MAX232 IC converts TTL signal from the microcontroller into RS232 voltage level [\ref{12}]. Since PIC12F675 has an internal oscillator of 4MHz, so it works without any external clock. General Purpose Input Output (GP I/O) pins GP0, GP1, GP2 and GP4 of IC3 were used as the analogue inputs. The MAX232 driver IC uses some external capacitors to enhance the voltage levels to RS232 level. A 9 Pin D Type female connector is used to enable communication with the COM port of the PC. The PCB of the circuit was designed using ExpressPCB.

\subsection{Sensors}

An IC temperature sensor LM35 (IC1) and a relative humidity sensor TPS-00715 (IC2) were used. LM35 is pre calibrated in Degree Celsius. The characteristic graph of the LM35 (Fig 2) is made from the experimental observation, which is linear. Also, the characteristic graph of the TPS-00715 (Fig 3) is made from the data sheet.

\subsection{Signal conditioning circuit}

Signal levels from the temperature sensor IC LM35 and humidity sensor IC TPS-00715 are not suitable for feeding it to the DAS directly.
\begin{figure}[!h]
\centering
\begin{minipage}[b]{0.4\linewidth}
\includegraphics[height = 5cm]{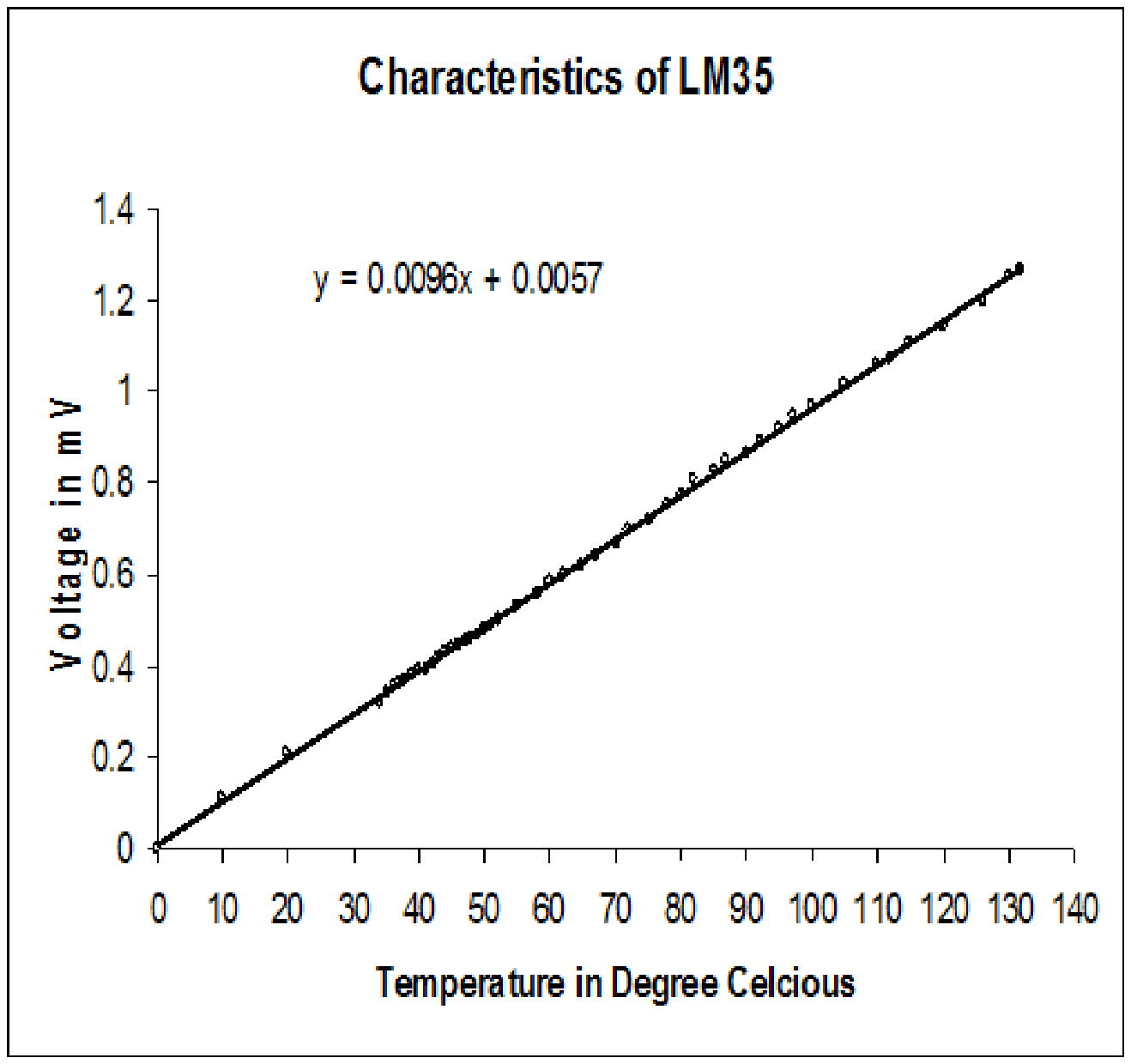}
\centering{\caption{Characteristics graph of the LM35}}
\end{minipage}%
\hspace{2 cm}
\begin{minipage}[b]{0.4\linewidth}
\includegraphics[height = 5cm]{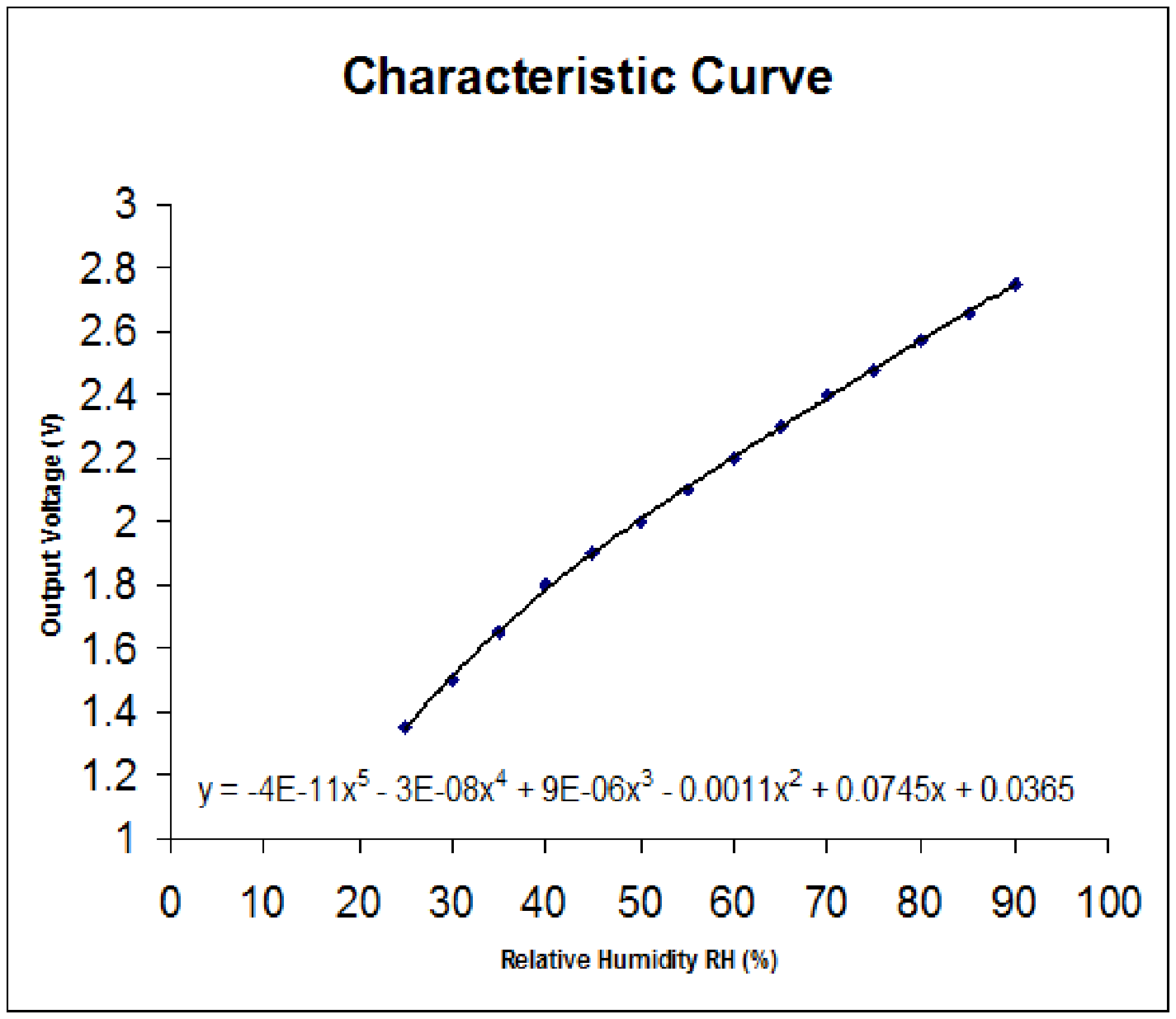}
\centering{\caption{Characteristics graph of the TPS-00715}}
\end{minipage}
\end{figure}
 For these, signal conditioning circuits have been designed using OpAmp OP07 as shown in Fig. 1. To protect the over voltage due to accidental or over range, a zener diode was used in series with a series resistor.  The output was taken across the zener diode, limiting to a mximum voltage of 5.1 volt, and fed to the input lines of the microcontroller.

\subsection{Software section}

For the proper functioning of the data acquisition system, a firmware was developed and written to the microcontroller and an application program was also developed in Visual Basic 6.

\subsubsection{Firmware}

A BASIC program was written in Oshonsoft PIC Simulator IDE for proper ADC conversion at a fixed sampling rate and sending the digitized data serially. The program was compiled to make a hex file. The hex file so generated was downloaded to the PIC12F675 microcontroller using PICkit2 programmer [\ref{13}]. The flowchart of the firmware program so developed is given in Fig. 4.
\begin{figure}[!h]
\centering \includegraphics[height=10 cm]{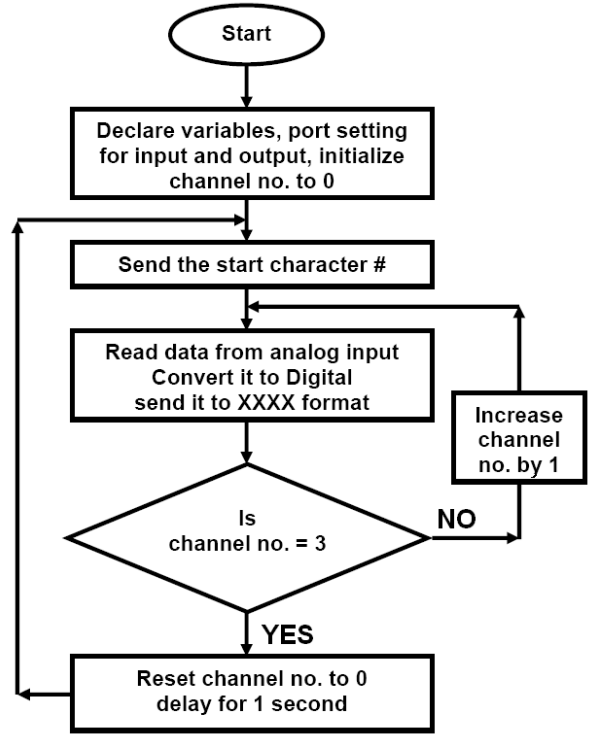}
\caption{Flow chart of the firmware}
\end{figure}
      
\subsubsection{Application program}

For proper acquisition of the data, graphical display into the monitor and saving it as a file into the hard disk of the PC, an application program was developed in Visual Studio 6.0. For preventing data missed, polling technique was used, that does not require a hardware interrupt [\ref{14}]. It uses the MSComm component to communicate through the comm port of the PC. In Visual Basic, MSComm’s OnComm event performs the event-driven routine which automatically jump to a routine when an event occurs. The application responds quickly and automatically to activity at the port, without having to waste time checking. The data so acquired is split into four values having four digit each, and analysed. Since it has 10 bit resolution it can read a value from 0 to 5 volt in 1023 steps for a channel. Thus, it has an accuracy of 4.88mV.\\
\begin{figure}[!h]
\centering \includegraphics[width=\linewidth]{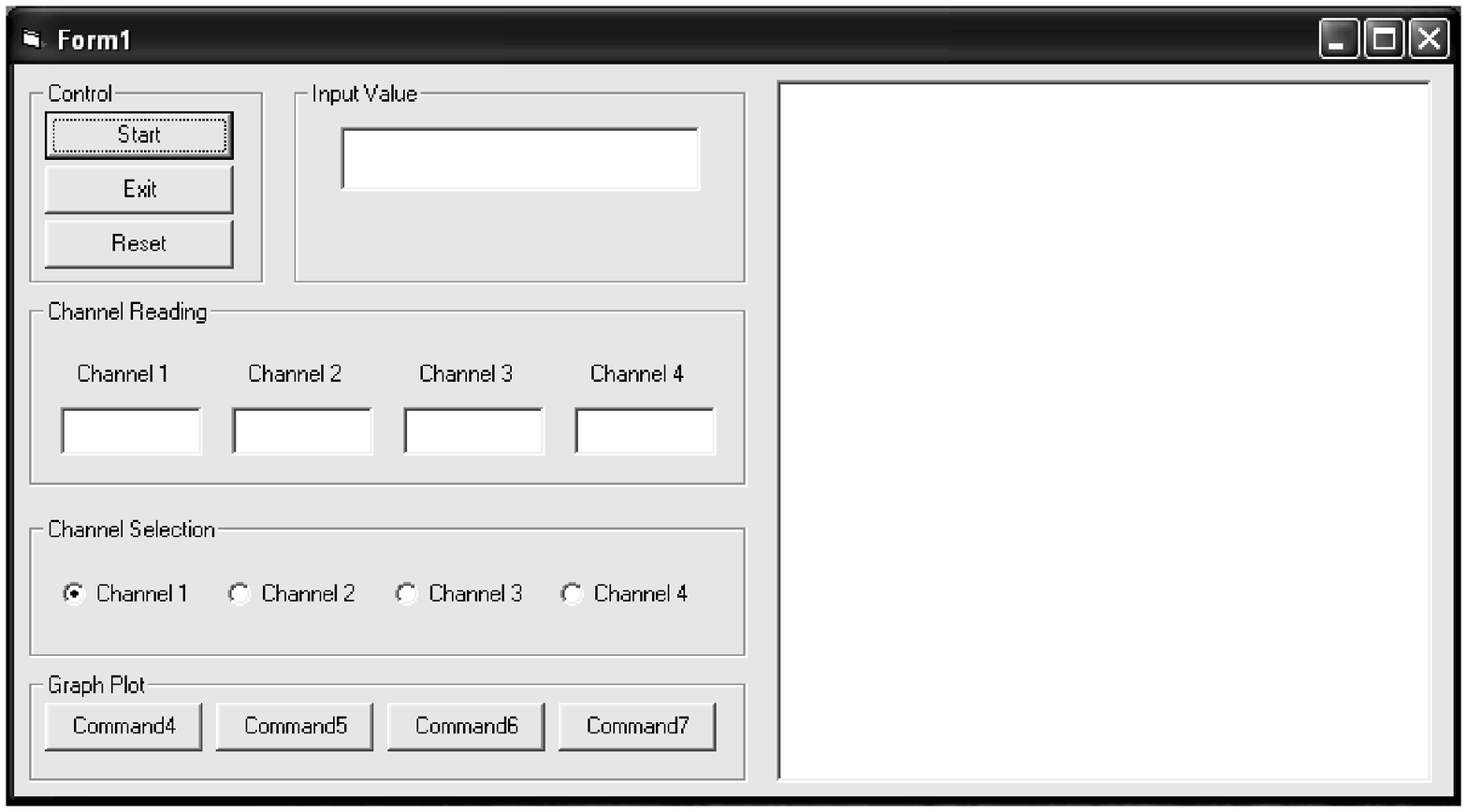}
\caption{Flow chart of the firmware}
\end{figure}

For the temperature, the input range is made from 0 to 50$^o$C. This is converted into voltage from 0 to 5 volt, by the signal conditioning circuit. The range of the relative humidity is from 10 to 90 \% RH having 1 to 3 volt output. Then, the signal conditioning circuit converted this voltage to a range from 1 to 5 volt. Since, the humidity sensor response is not linear, a polynomial fitting is made. The expression for 5th order polynomial fitting (as obtained from experimental data) is given by Eq. 1 as follows:\\
\begin{equation}
y = 15.538x^{5} - 161.37x^{4} + 655.54x^{3} - 1289.1x^{2} + 1259.3x - 472.15
\end{equation}

where x represents the voltage reading in volt  and y the relative humidity (RH) in percent (\%). This equation is fitted into the program of the application software, to give RH in percentage directly. The screen shot captured of the application program is shown in Fig. 5. The calibration curve of the humidity sensor TPS-00715 is made using polynomial fitting of 5th order using eqn. (1) as shown in Fig 6.

\subsection{Results and Conclusion}

According to our channel selection a graph is plotted for the selected channel. All the data are stored into the hard disk of the PC into comma separated value (.csv) format. The data so stored can be used for future analysis and also the wave form can be reconstructed. For temperature and humidity one representative curve for 12 hour continuous observation is presented. The obtained real time data for temperature and humidity is presented in the Fig. 7. RH1 and RH2 are the observed humidity from the two different digital thermo hygrometers Model No. IT-202, and RH3 is taken from the dry and wet bulb psychometric measurement. RH4 is observed by the designed system. All the four graph shows the similar nature of variations in humidity. The commercial digital hygrometers shows a difference of 9\% RH as seen from RH1 and RH2. The slight fluctuation in psychometric observation RH3 may be due to fluctuation in air flow and resolution of the thermometers’ reading. The variation between RH1 and RH4 is less than $\pm 2\%$, where 5\% is considerable in general, for RH measurement. Thus, the designed system gives better result.\\

\begin{figure}[!h]
\centering
\begin{minipage}[b]{0.4\linewidth}
\includegraphics[height = 5cm]{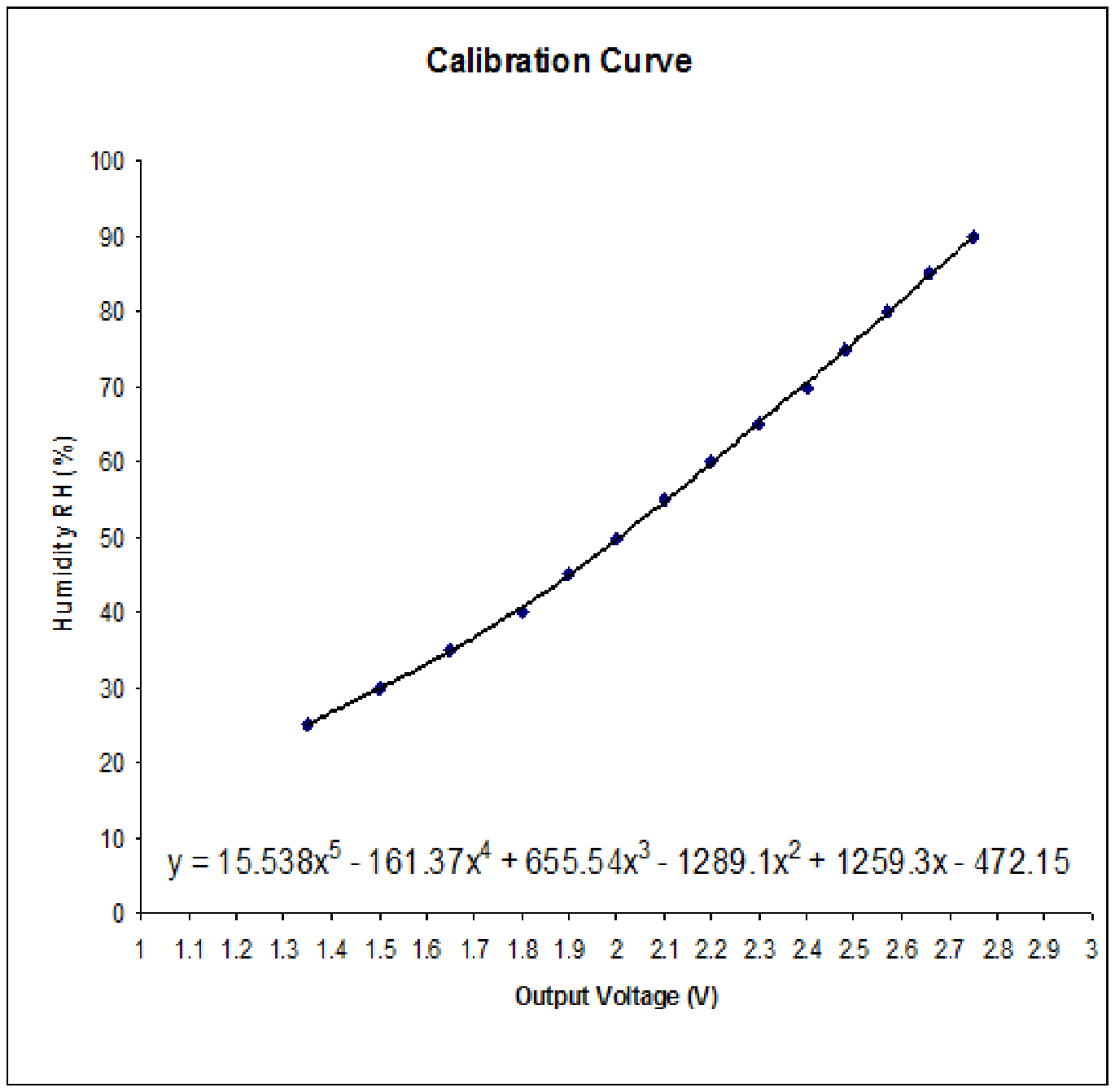}
\centering{\caption{Calibration graph of the TPS-00715}}
\end{minipage}
\hspace{2 cm}
\begin{minipage}[b]{0.4\linewidth}
\includegraphics[height = 5cm]{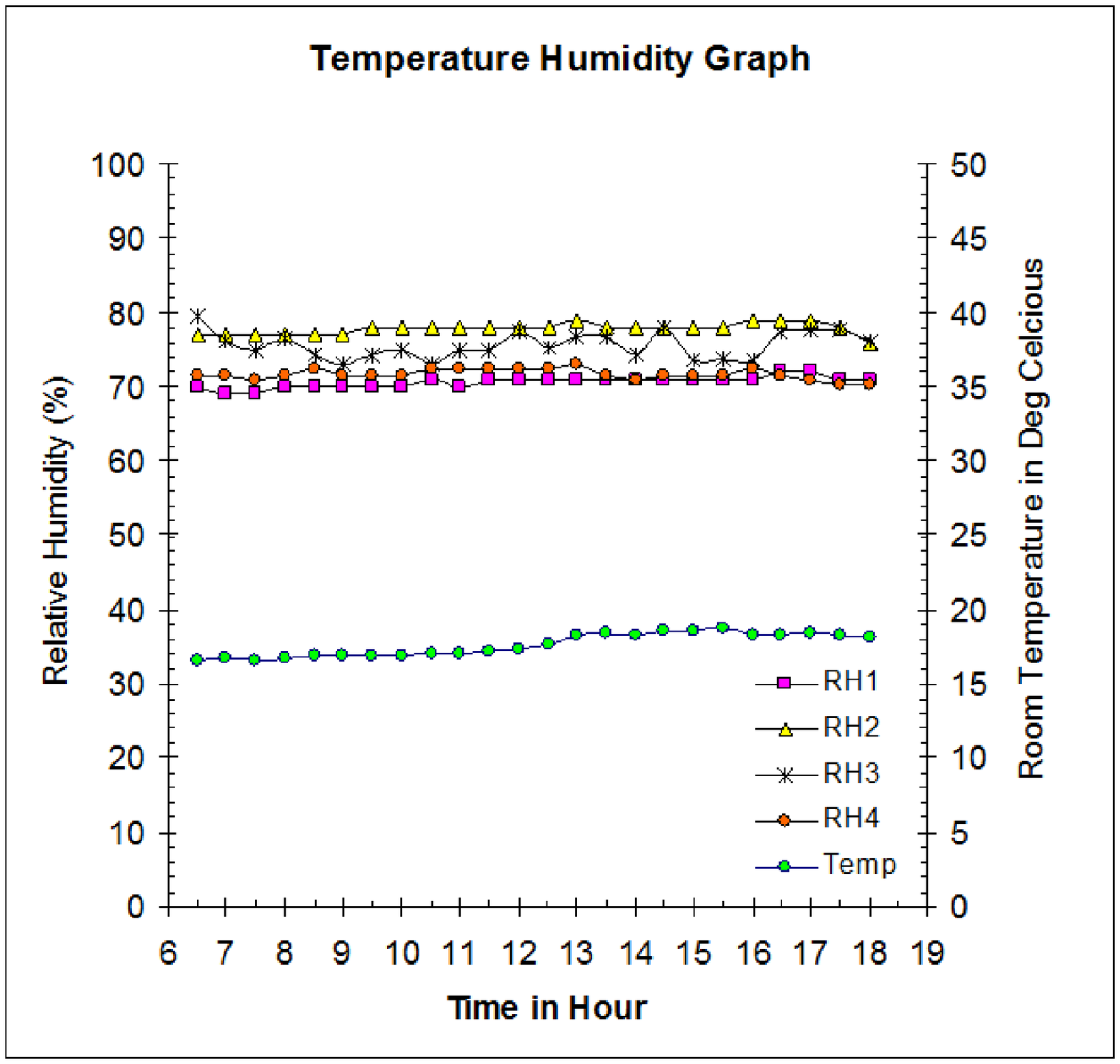}
\centering{\caption{Reconstructed waveform of the Temperature and Humidity}}
\end{minipage}
\end{figure}

This system will be useful in research and practical laboratories where acquisition for the measurement, monitoring, analysis and storage of temperature and relative humidity are necessary. In addition, the system can also be used in test and calibration laboratory. The designed system is a low cost with 10 bit resolution having accuracy of 4.88mV and compatible to PC and laptops. It can also be interfaced to the commonly available USB port of the PC or laptop using adapter USB to Serial adaptor BAFO BF-810 [\ref{15}]. Thus, the designed low cost device works with the legacy hardware as well as the modern USB interface.\\

\textbf{Acknowledgement:} The author (N. Monoranjan Singh) acknowledge the financial supports from the Department of Science \& Technology, New Delhi under the DST-INSPIRE Fellowship Scheme. The author also acknowledge Texas Instruments and Maxim-IC for providing help and support in the design and development.



\end{document}